\documentclass[sigconf]{acmart}

\usepackage{amsmath}
\usepackage{multirow}
\usepackage{bm}
\usepackage{bbm}
\usepackage[noend]{algpseudocode}
\usepackage{algorithmicx,algorithm}
\usepackage{subfigure}

\AtBeginDocument{%
  \providecommand\BibTeX{{%
    \normalfont B\kern-0.5em{\scshape i\kern-0.25em b}\kern-0.8em\TeX}}}

\setcopyright{acmcopyright}
\copyrightyear{2022}
\acmYear{2022}
\acmDOI{10.1145/1122445.1122456}

\acmConference[-]{-}{May 20, 2022}{-}
\acmBooktitle{CIKM' 2022}
\acmPrice{15.00}
\acmISBN{XXXXXX}

\begin{document}

\title{Sampling Is All You Need on Modeling Long-Term User Behaviors \\ for CTR Prediction}


\setcopyright{acmcopyright}
\copyrightyear{2022}
\acmYear{2022}
\acmDOI{10.1145/1122445.1122456}

\acmConference[CIKM '22]{CIKM '22: 31st ACM International Conference on Information and Knowledge Management}{October 17-21, 2022}{Hybrid Conference, Georgia, USA}
\acmBooktitle{CIKM '22: 31st ACM International Conference on Information and Knowledge Management, October 17-21, 2022, Georgia, USA}
\acmPrice{15.00}
\acmISBN{978-1-4503-XXXX-X/18/06}

\author{Yue Cao, XiaoJiang Zhou, Jiaqi Feng \\ Peihao Huang,  Yao Xiao, Dayao Chen, Sheng Chen}
\affiliation{%
  \institution{Meituan Inc.}
  \city{Beijing}
  \country{P.R. China}}
\email{yuecao@pku.edu.cn, zhouxiaojiang01@gmail.com, {fengjiaqi03, huangpeihao, xiaoyao06, chendayao, chensheng19}@meituan.com}

\renewcommand{\shortauthors}{Cao et al.}

\begin{abstract}
Rich user behavior data has been proven to be of great value for Click-Through Rate (CTR) prediction applications, especially in industrial recommender, search, or advertising systems. However, it's non-trivial for real-world systems to make full use of long-term user behaviors due to the strict requirements of online serving time. Most previous works adopt the retrieval-based strategy, where a small number of user behaviors are retrieved first for subsequent attention. However, the retrieval-based methods are sub-optimal and would cause information losses, and it's difficult to balance the effectiveness and efficiency of the retrieval algorithm.

In this paper, we propose \textbf{SDIM} (\textbf{S}ampling-based \textbf{D}eep \textbf{I}nterest \textbf{M}odeling), a simple yet effective sampling-based end-to-end approach for modeling long-term user behaviors. We sample from multiple hash functions to generate hash signatures of the candidate item and each item in the user behavior sequence, and obtain the user interest by directly gathering behavior items associated with the candidate item with the same hash signature. We show theoretically and experimentally that the proposed method performs on par with standard attention-based models on modeling long-term user behaviors, while being sizable times faster. We also introduce the deployment of SDIM in our system. Specifically, we decouple the behavior sequence hashing, which is the most time-consuming part, from the CTR model by designing a separate module named BSE (Behavior Sequence Encoding). BSE is latency-free for the CTR server, enabling us to model extremely long user behaviors. Both offline and online experiments are conducted to demonstrate the effectiveness of SDIM. SDIM now has been deployed online in the search system of Meituan APP. 
\end{abstract}




\maketitle

\section{Introduction}
Click-Through Rate (CTR) prediction is an essential task in industrial applications systems. User interest modeling, which aims at learning user's implicit interest from 
historical behavior data, has been widely introduced for real-world systems and contributes remarkable improvement \cite{DBLP:conf/kdd/ZhouZSFZMYJLG18, DBLP:conf/wsdm/TangW18, DBLP:conf/ijcai/FengLSWSZY19}. 

Various models are proposed for modeling users' interests \cite{DBLP:conf/wsdm/TangW18, DBLP:conf/icdm/KangM18, DBLP:conf/kdd/ZhouZSFZMYJLG18}. Among them, DIN \cite{DBLP:conf/kdd/ZhouZSFZMYJLG18} adaptively calculates user interests by taking the relevance of historical behaviors given a target item into consideration. DIN introduces a new attention module named \textbf{target attention}, where the target item acts as the query $\mathbf{Q}$ and historical user behaviors act as the key $\mathbf{K}$ and the value $\mathbf{V}$.
Due to its superior performance, DIN-based methods have become the mainstream solution for modeling user interests in recent years. However, the strict requirements of online serving time limit the length of user behavior sequences that can be used. As a result, most of the industrial systems truncate the user behavior sequences and just feed recent 50 behaviors \cite{DBLP:conf/kdd/ZhouZSFZMYJLG18, DBLP:conf/aaai/ZhouMFPBZZG19} for user interest modeling, which leads to information losses.

With the rapid development of the internet, user accumulates more and more behavior data on E-commerce platforms.
Take Taobao\footnote{The largest online shopping platform in China} for example, they report that 23\% of users have more than 1,000 behaviors in Taobao APP in six months \cite{DBLP:conf/sigir/RenQF0ZBZXYZG19}. In Meituan APP, there are more than 60\% of users have at least 1,000 behaviors, and more than 10\% of users have at least 5,000 behaviors in one year. How to effectively utilize more user behaviors for a more accurate user interest estimation becomes more and more important for industrial systems.


Recently, some methods are proposed to model users' long-term interests from long behavior sequences \cite{DBLP:conf/kdd/PiBZZG19, DBLP:conf/cikm/PiZZWRFZG20, DBLP:conf/sigir/Qin0WJF020, DBLP:journals/corr/abs-2108-04468}. MIMN \cite{DBLP:conf/kdd/PiBZZG19} decouples the user interest modeling from the entire model by designing a separate User Interest Center (UIC) module. Although UIC can save lots of online serving time, it makes it impossible for the CTR model to exploit the information from the target item, which has been proved to be crucial for users interest modeling \cite{DBLP:journals/corr/abs-2108-04468}. As a consequence, MIMN can only model shallow user interests.
SIM \cite{DBLP:conf/cikm/PiZZWRFZG20} and UBR4CTR \cite{DBLP:conf/sigir/Qin0WJF020} adopt two-stage frameworks for handling long-term user behaviors. They retrieves top-$k$ similar items from the sequence, and then feed these items to the subsequent attention module \cite{DBLP:conf/kdd/ZhouZSFZMYJLG18}. As pointed by \cite{DBLP:journals/corr/abs-2108-04468}, the retrieve objectives of these approaches are divergent with the goal of the CTR model, and the pre-trained embedding in offline inverted index is not suiting for online learning systems. 
To improve the quality of the retrieve algorithm, ETA \cite{DBLP:journals/corr/abs-2108-04468} proposes an LSH-based method to retrieve top-$k$ similar items from user behaviors in an end-to-end fashion. They use locality-sensitive hash (LSH) to convert items into hash signatures, and then retrieve top-$k$ similar items based on their hamming distance to the candidate item. The use of LSH greatly reduces the cost to calculate the similarity between items, and ETA achieves better results than SIM \cite{DBLP:conf/cikm/PiZZWRFZG20} and UBR4CTR \cite{DBLP:conf/sigir/Qin0WJF020}.

SIM, UBR4CTR, and ETA are all retrieval-based approaches. The retried-based methods suffer from following drawbacks: Retrieving top-$k$ items from the whole sequence is sub-optimal and would produce biased estimation of user's long-term interests. In the case that a user has rich behaviors, the retrieved top-$k$ items may all similar to the candidate item and the estimated user interest representation would be inaccurate.
Besides, it's difficult to balance to effectiveness and efficiency of the retrieval algorithm. Take SIM (hard) as an example, they use a simple retrieve algorithm, therefore its performance is inferior compared with other methods. In contrast, UBR4CTR achieves significant improvement with the aid of a complex retrieval module, but its inference speed becomes $4\times$ slower \cite{DBLP:conf/sigir/Qin0WJF020}, which prevents UBR4CTR from being deployed online, especially for long-term user behaviors modeling.  


In this paper, we propose a simple hash sampling-based approach for modeling long-term user behaviors.
First, we sample from multiple hash functions to generate hash signatures of the target item and each item in user behavior sequence. Instead of retrieving top-$k$ similar items using a certain metric, we directly gather behavior items that share the same hash signature with the target item from the whole sequence to form the user interest. 
The intrinsic idea behind our method is to approximate the softmax distribution of user interest with LSH collision probability. 
\textit{We show theoretically that this simple sampling-based method produces very similar attention patterns as softmax-based target attention and achieves consistent model performance, while being much more time-efficient}. As a consequence, our method behaviors like computing attention directly on the original long sequence, without information loss.
We named the proposed method as \textbf{SDIM} (means \textbf{S}ampling-based \textbf{D}eep \textbf{I}nterest \textbf{M}odeling). 

We also introduce our hands-on practice on deploying SDIM online. Specifically, we decouple our framework into two parts: (1) Behavior Sequence Hashing (BSE) server, and (2) CTR server, where the two parts are deployed separately. The behavior sequence hashing is the most time-consuming part of the whole algorithm, and the decoupling of this part greatly reduces the serving time. More details will be introduced in Section~\ref{dep}.

We conduct experiments on both public and industrial datasets. Experimental results show that SDIM achieves results consistent with standard attention-based methods, and outperforms all competitive baselines on modeling long-term user behaviors, and with sizable speed ups. SDIM has been deployed in the search system in Meituan\footnote{\url{https://meituan.com/}}, the largest e-commerce platform for lifestyle services in China, and bringing 2.98\% CTR and 2.69\% VBR lift, which is very significant to our business.

In sum, the main contributions of this paper are summarized as follows:
\begin{itemize}
\item We propose SDIM, a hash sampling-based framework for modeling long-term user behaviors for CTR prediction.
We prove that this simple sampling-based strategy produces very similar attention patterns as target attention.
\item We introduce our hands-on practice on deploying SDIM online in detail. We believe this work would help to advance the community, especially in modeling long-term user behaviors.
\item Extensive experiments are conducted on both public and industrial datasets, and the results demonstrate the superiority of SDIM in terms of efficiency and effectiveness. SDIM has been deployed in the search system in Meituan, contributing significant improvement to the business.
\end{itemize}

\section{Related Work}
\subsection{Click-Through Rate (CTR) Prediction}
With the rapid development of deep learning, deep neural-based methods have achieved remarkable performance in CTR prediction. 


User interest modeling is the key problem for CTR prediction. Many work \cite{DBLP:conf/kdd/ZhouZSFZMYJLG18, DBLP:conf/aaai/ZhouMFPBZZG19, DBLP:conf/ijcai/FengLSWSZY19, DBLP:conf/cikm/LiLWXZHKCLL19} focus on learning better representation from user historical behaviors in recent years. DIN \cite{DBLP:conf/kdd/ZhouZSFZMYJLG18} introduces target attention which learns different user interests with regard to different target items. In \cite{DBLP:conf/aaai/ZhouMFPBZZG19}, considering interest evolving process in user behaviors, DIEN proposes an interest evolving layer to capture dynamic interest about target item. In DSIN \cite{DBLP:conf/ijcai/FengLSWSZY19}, based on a prior knowledge that user behaviors are highly homogeneous in each session and heterogeneous in different sessions, the session-based model is proposed. 
It is worth noting that the target attention mechanism in DIN \cite{DBLP:conf/kdd/ZhouZSFZMYJLG18} has been widely used in CTR models nowadays.

\subsection{Long-Term Sequential User Behavior Modeling}
User behavior modeling has shown great performance in industrial applications. Exploring richer user behavior in CTR models has attracted much attention. However, long-term user behavior modeling is faced with several challenges, such as complex model deployment and system latency. 

In MIMN \cite{DBLP:conf/kdd/PiBZZG19}, the user interest center (UIC) block with a memory-based architecture design is proposed to tackle the challenge. The block is updated by user behavior event and only need to store limited user interest memory. MIMN is hard to model the interaction between user behavior sequence and target item which has proved to be important in CTR modeling. In SIM \cite{DBLP:conf/cikm/PiZZWRFZG20}, a two-stage method is proposed to model long-term user behavior sequences. Firstly, a general search unit (GSU) is utilized to extract relevant behaviors with regard to the target item. Secondly, an exact search unit is proposed to model relevant behaviors in an end-to-end manner. UBR4CTR \cite{DBLP:conf/sigir/Qin0WJF020} uses a similar search-based method as SIM \cite{DBLP:conf/cikm/PiZZWRFZG20} to face challenges. Recently, ETA \cite{DBLP:journals/corr/abs-2108-04468} proposes an end-to-end target attention method to model long-term user behavior sequence. They apply locality-sensitive hashing (LSH) to reduce the training and inference time cost.

Besides the above related works in the field of CTR prediction, there are also plenty of works in Natural Language Processing (NLP) that aim to improve the efficiency of self-attention \cite{DBLP:conf/iclr/KitaevKL20, DBLP:conf/aaai/ZhouZPZLXZ21, DBLP:conf/icml/ZengXRAFS21, DBLP:journals/corr/abs-2006-04768}. These approaches can reduce the time complexity of self-attention from $O(L^2)$ to $O(L\log(L))$, where $L$ is the sequence length, but can not be applied to reduce the $O(L)$ time complexity of the target attention.

\section{PRELIMINARIES}
\subsection{Task Formulation}
CTR prediction is a core task in the industrial recommendation, search, and advertising systems. The goal of the CTR task is to estimate the probability of a user clicking on the item, which is defined as follows:
\begin{equation}
    \text{prob}_i = P(y_i = 1 | x_i; \theta)
\end{equation}

In above equation, $\theta$ represents trainable parameters in CTR model. Given input features $x_i$, CTR model is trained to predict the click probability $\text{prob}_i$ by minimizing the cross-entropy loss: 
\begin{equation}
    L = -\frac{1}{N}\sum_{i=1}^{N}(y_i\log(\text{prob}_i)+(1-y_i)\log(1-\text{prob}_i))
\end{equation}

\subsection{Target Attention}
The concept of target attention is first proposed by DIN \cite{DBLP:conf/kdd/ZhouZSFZMYJLG18} and is widely applied for modeling user interests on CTR tasks \cite{DBLP:conf/aaai/ZhouMFPBZZG19, DBLP:conf/ijcai/FengLSWSZY19, DBLP:journals/corr/abs-2108-04468, DBLP:conf/cikm/PiZZWRFZG20, DBLP:conf/kdd/PiBZZG19}. Target attention takes the target item as query and each item in the user behavior sequence as key and value, and uses an attention operator to soft-search relevant parts from the user behavior sequence. The user interests are then obtained by taking a weighted sum over user behavior sequence. 

Specifically, designate the target item as $\mathbf{Q} \in \mathbb{R}^{B \times d}$ and user sequence representations as $\mathbf{S} \in \mathbb{R}^{B \times L \times d}$, where $B$ is the number of candidate items to be scored by the CTR model for each request, $L$ is the length of user behavior sequence, and $d$ is the model's hidden size.
Let $\mathbf{q}_i$ be the $i$-th target item, target attention calculates the dot product similarity between $\mathbf{q}_i$ and each item in behavior sequence $\mathbf{S}$, and then uses the normalized similarities as the weights to obtain user's interest representation. 
\begin{equation}
    \text{TA}(\mathbf{q}_i, \mathbf{S}) = \frac{\exp{(\mathbf{q}_i^\top \mathbf{s}_j / t)}}{\sum_{j=1}^{L}\exp{(\mathbf{q}_i^\top \mathbf{s}_j / t)}} \cdot \mathbf{s}_j \label{ta1}
\end{equation}
The matrix form of Eq.~\ref{ta1} can be written as:
\begin{equation}
    \text{TA}(\mathbf{Q}, \mathbf{S}) = softmax(\frac{\mathbf{Q}^\top \mathbf{S}}{t})\mathbf{S} \label{ta}
\end{equation}
The scaling factor $t$ is used to avoid large value of the inner product \cite{DBLP:journals/corr/abs-2108-04468}. 

The complexity of calculating $\text{TA}(\mathbf{Q}, \mathbf{S})$ is $O(BLd)$ where $B$ is the number of candidate items for each request, $L$ is the length of user behavior sequence, and $d$ is the model's hidden size. In large-scale industrial systems, $B$ is about $10^{3}$ and $d$ is about $10^2$, and \textbf{it's infeasible to deploy target attention on long-term user behaviors modeling for online systems} \cite{DBLP:journals/corr/abs-2108-04468, DBLP:conf/kdd/PiBZZG19}.

\subsection{Locality-Sensitive Hash (LSH) and SimHash} \label{sec.lsh}
Locality-sensitive hash (LSH) \cite{DBLP:conf/nips/AndoniILRS15} is an algorithmic technique for finding nearest neighbors efficiently in high-dimensional spaces. LSH satisfies the locality-preserving property: nearby vectors get the same hash with high probability while distant ones do not. Benefiting from this property, LSH-based signatures has been widely used in many applications such as web search system \cite{DBLP:conf/www/MankuJS07}. 
The \textbf{random projection scheme (SimHash)} \cite{DBLP:conf/stoc/Charikar02, DBLP:books/cu/LeskovecRU14} is an efficient implementation of LSH.
The basic idea of SimHash is to sample a random projection (defined by a normal unit vector $\mathbf{r}$) as hash function to hash input vectors into two axes (+1 or -1). 
Specifically, given an input $\mathbf{x}$ and a hash function $\mathbf{r}$, the corresponding hash is calculated as:
\begin{equation}
    h(\mathbf{x}, \mathbf{r}) = \text{sign}(\mathbf{r}^\top \mathbf{x}) \label{eq.lsh}
\end{equation} 
where $\mathbf{r} \in \mathbb{R}^d$, $r_i \sim \mathcal{N}(0,1)$. $h(\mathbf{x}, \mathbf{r}) = \pm 1$ depends on which side of the hashed output lies. For two vectors $\mathbf{x}_1$ and $\mathbf{x}_2$, we say $\mathbf{x}_1$ collides with $\mathbf{x}_2$ only when they have the same value of the hash code:
\begin{equation}
    p^{(\mathbf{r})} = \mathbbm{1}_{h(\mathbf{x}_1, \mathbf{r}) = h(\mathbf{x}_2, \mathbf{r})}
\end{equation}

While a single hash would work in estimating similarities, the outputs can be the average of multiple hashes to lower the estimation error. In practice, the $(m,\tau)$-parameterized SimHash (multi-round hash) algorithm \cite{DBLP:books/cu/LeskovecRU14} is usually adopted, where $m$ is the number of hash functions and $\tau$ is the width parameter. 
In the first step, SimHash randomly samples $m$ hash functions and generates $m$ hash codes for each input:
\begin{equation}
    h(\mathbf{x}, \mathbf{R}) = \text{sign}(\mathbf{R}\mathbf{x}) \label{eq.simhash}
\end{equation} 
where $\mathbf{R} \in \mathbb{R}^{m \times d}$, $h(\mathbf{x}, \mathbf{R}) \in \mathbb{R}^{m}$. 
To reduce the probability of dissimilar items having the same hash codes, the algorithm merge the results of every $\tau$ codes to form a new hash signature. For more details, please refer to \cite{DBLP:books/cu/LeskovecRU14}. In SimHash, $\mathbf{x}_1$ collides with $\mathbf{x}_2$ only when they have the same value of the hash signature, i.e., their hash codes in a batch of $\tau$ codes are all the same:  
\begin{equation}
\begin{aligned}
    \tilde{p}^{(\mathbf{R})} &= p^{(\mathbf{r}_{1})}\ \& \  p^{(\mathbf{r}_{2})}\ \& \ \cdots \& \ p^{(\mathbf{r}_{\tau})} \\
    &= \&_{k=1}^{\tau} p^{(\mathbf{r}_{k})}  
\end{aligned}
\end{equation}
where "$\&$" represents the operator of logical "AND", and $\mathbf{r}_{k}$ is the $k$-th hash function. The bottom of Figure~\ref{fig1} shows an example of $(4,2)$-parameterized SimHash algorithm, where we use 4 hash functions and aggregate every 2 hash codes into 2 hash signatures (yellow and green).


\section{Methodology}
\begin{figure*}[tb]
	\centering 
	\includegraphics[width=0.9\linewidth]{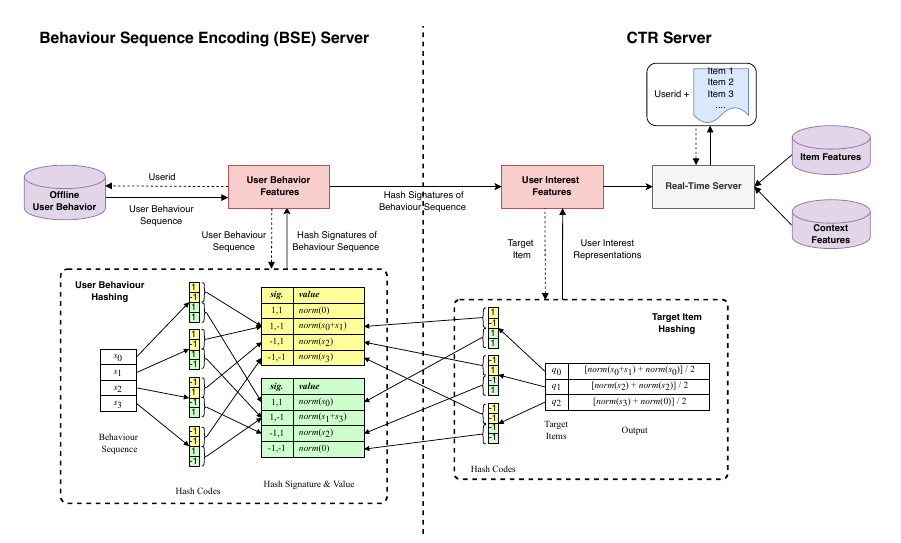}  
	\caption{High-level overview of our framework. It consists of two separate servers: Behavior Sequence Encoding (BSE) server and CTR server. The BSE server calculates the hashing of user behavior sequence, and the CTR server calculates the hashing of candidate items and gathers behavior items that share the same hash signature with the candidate item to form user interests. The CTR server contains a real-time CTR model, which takes user interest features, item features, and context features as inputs to predict the click probability corresponding to each candidate item.}  
	\label{fig1}   
\end{figure*}

In this section, we introduce our framework for implementing SDIM in our system. A high-level overview can be seen in Figure~\ref{fig1}. The framework is consist of two separate servers: Behavior Sequence Encoding (BSE) server and CTR Server, which will be introduced in detail later.

\subsection{User Behavior Modeling via Hash-Based Sampling}
\subsubsection{Hash Sampling-Based Attention}
At the first step, we sample multiple hash functions and generate the hash codes of user behaviors and candidate items. Similar to ETA \cite{DBLP:journals/corr/abs-2108-04468}, we use fixed random hash vectors drawn from normal distribution as "hash functions".
After hashing, ETA calculates the Hamming distance between items as an approximation of user interest distribution to select top-$k$ similar items, here we propose a more efficient and more effective way to approximate the user's interest.

With the locality-preserving property, similar vectors fall in the same hash bucket with high probability, therefore the similarity between user behavior items and the candidate item can be approximated by their frequency of having the same hash codes (signatures), or the collision probability. This leads us to hypothesize that \textbf{the probability of hash collision can be an effective estimator of user's interests.} 

Based on this observation, we propose a new method to obtain the user's interest with LSH. After hashing, we directly form the user interest by summing together behavior items $\mathbf{s}_j$ associated with the candidate item $\mathbf{q}$ with the same signature.
In a single hash function $\mathbf{r}$, the proposed method for estimating user interest can be calculated by:
\begin{equation}
    \ell_{2}\left(\mathbf{P}^{(\mathbf{r})} \mathbf{S}\right) = \ell_{2}\left(\sum_{j=1}^{L} p^{(\mathbf{r})}_{j} \mathbf{s}_{j}\right) \label{eq9}
\end{equation}
where $\mathbf{S}$ is the user behavior sequence and $\mathbf{s}_j$ is the $j$-th item in this sequence, $p^{(\mathbf{r})}_{j} = \{0,1\}$ specifies whether $\mathbf{s}_j$ contributes to the user interests. Concretely, if $\mathbf{s}_j$ shares the same hash signature with the candidate item $\mathbf{q}$ under hash function $\mathbf{r}$, then $p^{(\mathbf{r})}_{j} = 1$, else $p^{(\mathbf{r})}_{j} = 0$:
\begin{equation}
    p^{(\mathbf{r})}_{j} = \mathbbm{1}_{h(\mathbf{q}, \mathbf{r}) = h(\mathbf{s}_j, \mathbf{r})}
\end{equation}
where $h(\cdot,\cdot)$ is defined in Eq.~\ref{eq.lsh}. $\ell_{2}(\cdot)$ in Eq.~\ref{eq9} refers to the $\ell_{2}$ normalization, which is used to normalize the interest distribution such that the attention weights sum up to 1.\footnote{We also tried to normalize the distribution using the sum of $P^{(\mathbf{r})}_{j}$, and the resulting model performs on par with $\ell_{2}$ normalized model. However, the implementation of $\ell_{2}$ normalization will be more efficient, therefore we use the $\ell_{2}$ normalization.}

\subsubsection{Multi-Round Hashing}
There is always a small probability that dissimilar items share the same hash codes with the candidate item, thus bringing noise to the model. To reduce this probability, we use the multi-round hash algorithm described in Section~\ref{sec.lsh}.

Specifically, we use the $(m,\tau)$-parameterized SimHash algorithm. We sample and perform $m$ times hashes in parallel, and merge the results of every $\tau$ hash codes to form a new hash signature. We regard $\mathbf{s}_j$ collides with $\mathbf{q}$ only when they have the same value of aggregated hash signature, i.e., the codes of $\mathbf{s}_j$ within a batch of signature should be all the same with $\mathbf{q}$'s.
\begin{equation}
    \tilde{p}^{(\mathbf{R}_{i})}_{j} = \&_{k=1}^{\tau} p^{(\mathbf{r}_{i,k})}_{j}, \quad \text{where}\ p^{(\mathbf{r}_{i,k})}_{j} = \mathbbm{1}_{h(\mathbf{q}, \mathbf{r}_{i,k}) = h(\mathbf{s}_j, \mathbf{r}_{i,k})}
\end{equation}

The outputs of $m / \tau$ hash signatures are averaged for a low-variance estimation of user interest. It can be formulated as:
\begin{equation}
\begin{aligned}
    \text{Attn}(\mathbf{q}, \mathbf{S}) &= \frac{1}{m/ \tau} \sum_{i=1}^{m/ \tau} \ell_{2}(\tilde{\mathbf{P}}^{(\mathbf{R}_i)} \mathbf{S}) \\
    &= \frac{1}{m/ \tau} \sum_{i=1}^{m/ \tau} \ell_{2}(\sum_{j=1}^{L} \tilde{p}^{(\mathbf{R}_i)}_{j} \mathbf{s}_{j}) \label{attn}
\end{aligned}
\end{equation}

\subsection{Analysis of Attention Patterns}
\subsubsection{Expectation of Hash Sampling-Based Attention}
In our method, as $\mathbf{s}_j$ becomes more similar to $\mathbf{q}$, their collision probability becomes higher and the expectation of the coefficient $\tilde{p}_j$ is higher.
It can prove that the expectation of $\tilde{p}_{j}$ is the angle between $\mathbf{s}_j$ and $\mathbf{q}$ on the unit circle \cite{DBLP:conf/stoc/Charikar02}\footnote{We normalize $\mathbf{q}$ and $\mathbf{s}_j$ before calculating user interests.}:
\begin{equation}
    \mathbb{E}\left[\tilde{p}_{j} \right]=\left(1-\frac{\arccos \left(\mathbf{q}^{\top} \mathbf{s}_j\right)}{\pi}\right)^{\tau} \label{epj}
\end{equation}
Therefore, the expectation of user interest representation produced by SDIM is:
\begin{equation}
\begin{aligned}
    \mathbb{E}\left[\text{Attn}(\mathbf{q}, \mathbf{S}) \right]&= \mathbb{E}\left[ \ell_{2}(\sum_{j=1}^{L} \tilde{p}^{(\mathbf{R}_k)}_{j} \mathbf{s}_{j})\right] \\
    &= \frac{(1 - \frac{\arccos (\mathbf{q}^{\top} \mathbf{s}_j)}{\pi})^\tau}{\sum_{j=1}^{L} (1 - \frac{\arccos (\mathbf{q}^{\top} \mathbf{s}_j)}{\pi})^\tau} \cdot \mathbf{s}_j \label{EAttn}
\end{aligned}
\end{equation}

As the number of hash signatures $m/ \tau$ increases, $\tilde{p}_{j}$ converges to $\mathbb{E}\left[\tilde{p}_{j} \right]$, and $\text{Attn}(\mathbf{q}, \mathbf{S})$ converges to $\mathbb{E}\left[\text{Attn}(\mathbf{q}, \mathbf{S}) \right]$. 
The empirical results show that when $m/ \tau \geq 16$, the estimation error will be very small. In practice, we use $m=48$ and $\tau=3$ for our online model.

\begin{figure}[tb]
	\centering 
	\includegraphics[width=1.0\linewidth]{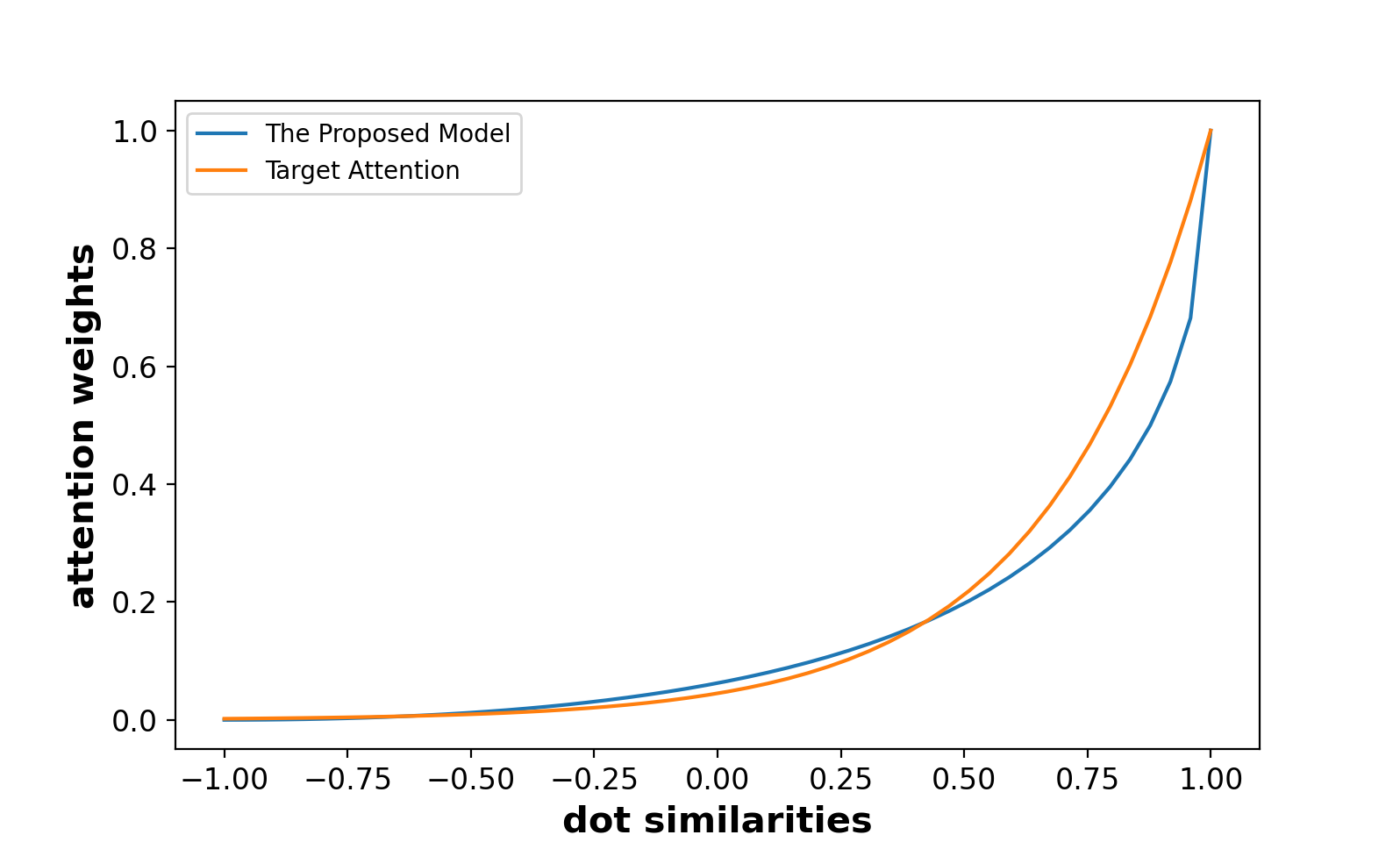}  
	\caption{Visualization of attention weights produced by target attention and our model. We left shift the curve of target attention by one unit ($\exp((x-1)/\sqrt{d})$) such that the input is in $[-1, 1]$ (Since the weights will be normalize, this will not change the attention weights). It can be seen that the proposed model produces very similar attention patterns as the target attention.}  
	\label{fig2}   
\end{figure}

We plot the attention weights $\left((1 - \frac{\arccos (\mathbf{q}^{\top} \mathbf{s}_j)}{\pi})^3\right)$ produced by SDIM in Figure~\ref{fig2}. For comparison, we also plot the attention weights $\left(\exp(\mathbf{q}^{\top} \mathbf{s}_j/0.5)\right)$ produced by target attention in the same figure. 
From Figure~\ref{fig2}, we can see that the attention weights of SDIM align well with the softmax function in target attention, therefore, \textbf{theoretically, our method can obtain outputs that are very close to the target attention}.
We name the proposed method as SDIM, which means \textbf{S}ampling-based \textbf{D}eep \textbf{I}nterest \textbf{M}odeling.

\subsubsection{The Property of $\tau$ and Relation to Other Methods} \label{sec-4.2.2}
In this subsection, we describe that the width parameter $\tau$ in SDIM plays a role in controlling the strength of the model on paying more attention to more similar items. 

Denote $w_j$ as the attention weight of target item $\mathbf{q}$ to item $\mathbf{s}_j$, i.e.,
\begin{equation}
    w_j = \frac{(1 - \frac{\arccos (\mathbf{q}^{\top} \mathbf{s}_j)}{\pi})^\tau}{\sum_{j=1}^{L} (1 - \frac{\arccos (\mathbf{q}^{\top} \mathbf{s}_j)}{\pi})^\tau} \label{wj}
\end{equation}
As $\tau$ increases, the entropy of the attention distribution $H(w_j)$ decreases strictly, and the distribution of $\mathbf{w}$ becomes sharper on large similarity region, which encourages the model to pay more attention to more similar items. 

Let us consider two extreme cases:
\begin{itemize}
\item When $\tau \rightarrow +\infty$ (can be approximated by assigning a large value), the algorithm only attends to items that are exactly the same as the candidate item. If we use the category attributes for hashing, then the algorithm behaviors like \textbf{SIM (hard)} \cite{DBLP:conf/cikm/PiZZWRFZG20}. Therefore, our algorithm can be seen as an extension of SIM (hard), where it will also considers very similar behavior items but with different category ids.
\item When $\tau = 0$, or $m=1$, the algorithm degenerates into \textbf{Mean Pooling} as the target item and user behavior items will always have the same hash signature. 
\end{itemize}

Therefore, SDIM is very flexible that can model different attention patterns by assigning different value of $\tau$.

\begin{figure*}[bt]
	\centering 
	\includegraphics[width=0.85\linewidth]{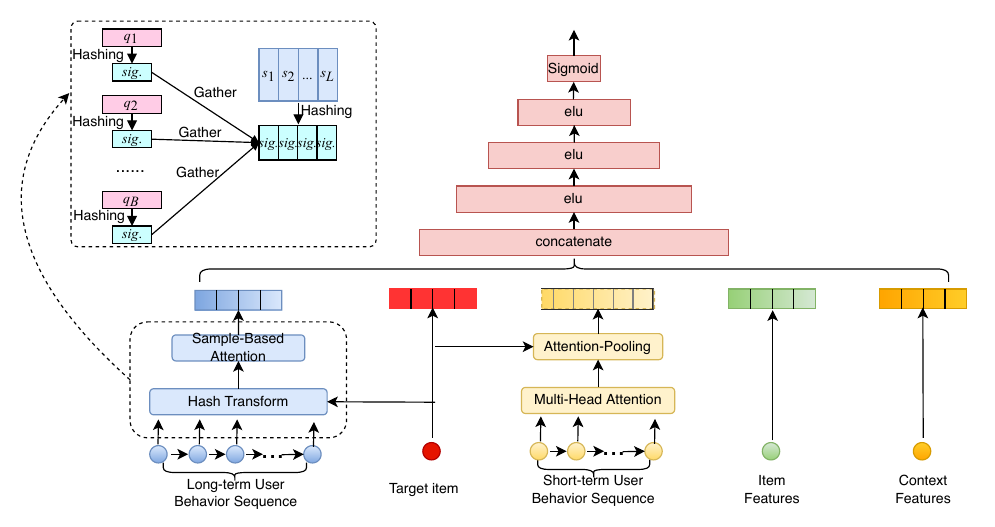}  
	\caption{Illustration of our online CTR model. The model takes user interest features, item features and context features as inputs to predict the click probability corresponds to each candidate item.}  
	\label{fig4}   
\end{figure*}

\subsection{Implementation and Complexity Analysis} \label{impl}

In this subsection, we elaborate that SDIM is sizable times faster than standard attention-based methods.

Let's review the formulation for calculating user interest in target attention (Eq.~\ref{ta}). This algorithm first gets the attention weights by multiplying the target vector with the behavior sequence, followed by a weighted sum of the behavior sequence using the attention weights.
Therefore, two matrix multiplications are needed for the representation of user interest in target attention, and the time complexity of this algorithm is $O(BLd)$.

SDIM decouples the calculation of user interest into two parts: (1) behavior sequence hashing, and (2) target vector hashing. \textbf{Note that user behavior sequence are users' inherent features and is independent of the candidate item, which means that the results of users' behavior sequence hashing are fixed no matter what the candidate item is.}
Therefore for each user, \textbf{we only need to compute the hash transform of user behavior sequence once in each request}. As a result, SDIM reduces $B$ to 1 in the time complexity, thus is sizable times faster than the standard target attention.
After hashing, sequence items that share the same hash signature with the target item are selected and summed together to form the user interest. 
In Tensorflow, this step can be implemented via a $\mathbf{tf.batch\_gather}$ operator, which is an atomic operator of Tensorflow and the time cost is negligible. 

The most time-consuming part of SDIM is the multi-round hashing of the behavior sequence, i.e., transform a $d$-dimensional matrix into $m$-dimensional hash codes. The time complexity of this operation is $O(Lmd)$ and can be reduced to $O(Lm\log(d))$ using the Approximating Random Projection algorithm \cite{DBLP:conf/nips/AndoniILRS15}. 
Since $m \ll B$ and $\log(d) \ll d$, \textbf{SDIM is much faster than standard attention-based methods}. In our scenario, SDIM achieves \textbf{10x-20x speedups} in training the user behavior modeling part.

\begin{table}[t!]
\begin{center}
\setlength{\tabcolsep}{2.0mm}{
\begin{tabular}{l|ccc}
\toprule \bf Method & \bf Training & \bf Online-Serving \\  \midrule
DIN & $O(BLd)$ & $O(BLd)$ \\
SIM & $O(B\log(M)+Bkd)$ & $O(B\log(M)+Bkd)$ \\
ETA & $O(Lm\log(d)+BLm+Bkd)$ & $O(BLm+Bkd)$ \\ 
SDIM & $O(Lm\log(d)+Bm\log(d))$ & $O(Bm\log(d))$ \\ \bottomrule
\end{tabular}}
\end{center}
\caption{\label{cplx} Time complexity of different methods at training and online-serving stage. $B$ is the number of candidate items for each request, $m$ is the number of hashes, $L$ and $k$ are the length of original and retrieved user behavior sequence respectively, $M$ is the size of attribute inverted index in SIM, and $d$ is the model's hidden size. In our system, $B$ is about $10^{3}$, $L=1024$, $m=48$, and $d=128$.}
\end{table}

\subsection{Deployment of the Whole System} \label{dep}
We introduce how we successfully deploy SDIM online in this subsection.
As described above, the whole algorithm is decoupled into two parts: (1) behavior sequence hashing, and (2) target vector hashing. The behavior sequence hashing is independent of the candidate item, which motivates us to build a specialized server to maintain user-wise behavioral sequence hashes.


We divide our system into two parts: \textbf{Behavior Sequence Encoding (BSE) server} and \textbf{CTR Server}, as shown in Figure~\ref{fig1}.
The BSE server is responsible for maintaining user-wise behavior sequence hashes. When received a list of user behaviors, BSE server samples from multiple hash functions and generates hash signatures for each item in behavior sequence. The items are then assigned to different buckets according to their signatures, where each bucket corresponds to a hash signature, as shown in Figure~\ref{fig1}. The hash buckets are passed to the CTR Server to model candidate item-aware user interests.

The CTR Server is responsible for predicting the click probability of candidate items. When received a batch of $B$ candidate items, the CTR Server hashes each item into signatures, and gathers item representations from the corresponding buckets.
The user interest features concatenated with other features are fed to a complex CTR model to get the predicted probabilities of each item. The overall structure of our CTR model is shown in Figure~\ref{fig4}. The model takes item features, context features, short-term user interest features, and long-term user interest features as inputs, and uses a multi-layer perceptron to predict the click probability of candidate items. 
Please note that SDIM does not require changing the structure of the CTR model, which can be naturally plugged into existing popular CTR architectures.
The proposed framework is \textbf{end-to-end} at the training stage: the user interest modeling part is jointly trained with the rest of the CTR model, and we deploy them separately only at the online serving stage. 

After decoupling BSE and CTR Server, the computation of the BSE is \textbf{latency-free} for the CTR Server. In practice, We put the BSE Server before the CTR Server and compute it in parallel with the retrieval module.

After deploying separately, the time cost of calculating user's long-term interest only lies in the hashing of candidate items, which has a $O(Bm\log(d))$ time complexity and is independent of the sequence length $L$, which means that our framework can handle the user interest modeling with extremely long behaviors theoretically. 
From the perspective of CTR Server, this time complexity is just feel like adding a common feature. In Table~\ref{cplx}, we compare the time complexity of different methods\footnote{Since UBR4CTR uses a complex neural network to select features, its time complexity cannot be accurately estimated.}.

Our serving system is somewhat similar to MIMN \cite{DBLP:conf/kdd/PiBZZG19}. The biggest difference is that our system can model users' deep interests, while MIMN can only model shallow interests.  

\subsubsection{Remark on Transmission Cost of Servers}
For each request, we need to transmit bucket representations from the BSE server to the CTR server. Notice that we use a fixed number of hash functions, therefore no matter how long the user's behavior is, we only need to transmit fixed-length vectors of bucket representations to the CTR server.
In our online system, the size of this vector is 8KB and the transmission cost is about 1ms.  

\section{Experiments}
\subsection{Dataset and Evaluation Metrics}
To verify the effectiveness of SDIM, we conduct experiments on both public and real-world industrial datasets.
For public dataset, we follow previous work \cite{DBLP:conf/kdd/PiBZZG19, DBLP:conf/cikm/PiZZWRFZG20, DBLP:journals/corr/abs-2108-04468} to choose Taobao dataset\footnote{\url{https://tianchi.aliyun.com/dataset/dataDetail?dataId=649}}. For industrial dataset, we use real-world data collected from the Meituan search system for experiments.

\textbf{Taobao Dataset:} Taobao dataset is released by \cite{DBLP:conf/kdd/ZhuLZLHLG18} and is widely used for offline experiments in previous work \cite{DBLP:conf/sigir/Qin0WJF020, DBLP:conf/cikm/PiZZWRFZG20}. Each instance in this dataset is consist of five fields of features: user ID, item ID, category ID, behavior type, and timestamp. Following \cite{DBLP:conf/sigir/Qin0WJF020}, we additionally introduce the feature of "is\_weekend" according to the timestamp to enrich context features. 
We pre-process the data in the same way with MIMN \cite{DBLP:conf/kdd/PiBZZG19} and ETA \cite{DBLP:journals/corr/abs-2108-04468}. Concretely, we use the  $1$st to $(L-1)$-th behaviors as inputs to predict the $L$-th behavior. We split the samples into training set (80\%), validation set (10\%), and test set (10\%) according to the timestep. The recent 16 behaviors are selected as short-term sequence, and the recent 256 behaviors are selected as long-term sequence.

\textbf{Industrial Dataset:} This dataset is collected from the platform searching system of Mobile Meituan APP, which is the largest e-commerce platform for lifestyle services in China. 
We select consecutive 14-day samples for training and the next two days for evaluation and testing, the number of training examples is about 10 billion. The recent 50 behaviors are selected as short-term sequence, and the recent 1,024 behaviors are selected as long-term sequence. If the number of user behaviors doesn't reach  this length, then we pad the sequence to the maximum length with the default value. Besides the user behavior features, we additionally introduce about 20 important id features to enrich the inputs. 

\textbf{Evaluation Metrics:}
For offline experiments, we follow previous work to adopt the widely used \textbf{Area Under Curve (AUC)} for evaluation.  
We also use the \textbf{Training \& Inference Speed (T\&I Speed)} as a supplement metric to show the efficiency of each model.
For online experiments, we use \textbf{CTR (Click-Through Rate)} and \textbf{VBR (Visited-Buy Rate)} as online metrics.

\subsection{Competitive Models}
Following previous work \cite{DBLP:journals/corr/abs-2108-04468, DBLP:conf/cikm/PiZZWRFZG20}, we compare SDIM with the following mainstream industrial models for modeling long-term user behaviors:
\begin{itemize}
\item \textbf{DIN} \cite{DBLP:conf/kdd/ZhouZSFZMYJLG18}. DIN is one of the most popular models for modeling user interest in industrial systems. However, DIN is infeasible to be deployed on modeling long-term user behaviors due to its high time complexity. 
In this baseline, we only use the short-term user behavior features but not long-term features.

\item \textbf{DIN (Long Seq.)}. For offline experiments, to measure the information gain of long-term user behavior sequences, we equip DIN with long behavior sequences. We set $L=256$ for Taobao dataset and $L=1024$ for industrial dataset.

\item \textbf{DIN (Avg-Pooling Long Seq.)}. This baseline is introduced by \cite{DBLP:journals/corr/abs-2108-04468} and \cite{DBLP:conf/cikm/PiZZWRFZG20}, where DIN is applied to modeling short-term user interest, and the long-term user interest is obtained by a mean pooling operation on long-term behaviors. We denote this baseline as \textbf{DIN(Avg-Pooling)}.

\item \textbf{SIM} \cite{DBLP:conf/cikm/PiZZWRFZG20}. SIM first retrieves top-$k$ similar items from the whole sequence via category id, and then applies target attention on top-$k$ items to get user interest. We follow previous work to compare SIM(hard) as the performance are almost the same they deploy SIM(hard) online.

\item \textbf{UBR4CTR} \cite{DBLP:conf/sigir/Qin0WJF020}. UBR4CTR is a two-stage method. At the first stage, they design a feature selection module to select features to form the query, and store the user behaviors in an inverted index manner. At the second stage, the retrieved behaviors are fed to a target attention-based module to get user interest.

\item \textbf{ETA} \cite{DBLP:journals/corr/abs-2108-04468}. ETA applies LSH to encode the target item and user behavior sequence into binary codes, and then computes the item-wise hamming distance to select top-$k$ similar items for subsequent target attention.
\end{itemize}

\textbf{MIMN} \cite{DBLP:conf/kdd/PiBZZG19} is proposed by the same team with SIM. As the authors claim that SIM defeats MIMN and they deploy SIM online, we only compare with SIM and omit the baseline of MIMN for the space limitation.

For all the baselines and SDIM, we use the same features (include timeinfo features) as input and adopt the same model structure except for the long-term user behavior modeling module. All models use the same length of long-term user behaviors (T=256 for Taobao and T=1024 for industrial dataset).


\section{Results and Analysis}
\subsection{Results on Taobao Dataset}
\begin{table}[t!]
\begin{center}
\setlength{\tabcolsep}{4.0mm}{
\begin{tabular}{l|cc}
\toprule \bf Method & \bf AUC $\uparrow$ & \bf T\&I Speed $\uparrow$ \\ \midrule
{DIN(Long Seq.)} \cite{DBLP:conf/kdd/ZhouZSFZMYJLG18} & 0.8848 & - \\
{DIN} \cite{DBLP:conf/kdd/ZhouZSFZMYJLG18} & 0.8627* & 7.4x \\ \hline
{DIN(Avg-Pooling)} \cite{DBLP:journals/corr/abs-2108-04468, DBLP:conf/cikm/PiZZWRFZG20} & 0.8669* & 2.6x \\
{SIM} \cite{DBLP:conf/cikm/PiZZWRFZG20} & 0.8692* & 2.4x \\ 
{UBR4CTR} \cite{DBLP:conf/sigir/Qin0WJF020} & 0.8752* & 0.8x \\ 
{ETA} \cite{DBLP:journals/corr/abs-2108-04468} & 0.8753* & 1.8x \\ 
{SDIM} & \bf 0.8854 & 5.0x \\
\bottomrule
\end{tabular}}
\end{center}
\caption{\label{result:1} Performance comparison of different model on Taobao dataset. The Training-Speed improvements are calculated on the basis of DIN(Long Seq.). "*" indicates that the improvement of SDIM over this baseline is statistically significant at p-value < 0.05 over paired Wilcoxon-test.}
\end{table}

\begin{table}[t!]
\begin{center}
\setlength{\tabcolsep}{4.0mm}{
\begin{tabular}{l|cc}
\toprule \bf Method & \bf AUC & \bf T\&I Speed \\ \midrule
{DIN(Long Seq.)} \cite{DBLP:conf/kdd/ZhouZSFZMYJLG18} & \bf 0.7049 & - \\
{DIN} \cite{DBLP:conf/kdd/ZhouZSFZMYJLG18} & 0.6652* & 13.5x \\ \hline
{DIN(Avg-Pooling)} \cite{DBLP:journals/corr/abs-2108-04468, DBLP:conf/cikm/PiZZWRFZG20} & 0.6749* & 11.0x \\
{SIM} \cite{DBLP:conf/cikm/PiZZWRFZG20} & 0.6852* & 10.8x \\ 
{UBR4CTR} \cite{DBLP:conf/sigir/Qin0WJF020} & 0.6836* & 1.2x \\ 
{ETA} \cite{DBLP:journals/corr/abs-2108-04468} & 0.6906* & 3.7x \\ 
{SDIM} & 0.7044 & 11.4x \\
\bottomrule
\end{tabular}}
\end{center}
\caption{\label{result:2} Performance comparison of different model on Industrial dataset. The Training-Speed improvements are calculated on the basis of DIN(Long Seq.). "*" indicates that the improvement of SDIM over this baseline is statiscally significant at p-value < 0.05 over paired Wilcoxon-test.}
\end{table}

The overall results of different models on Taobao dataset are shown in Table~\ref{result:1}. 
We can draw the following conclusions: (1) SDIM performs on par with DIN(Long Seq.) on modeling long-term user behaviors, while being 5x times faster. As described above, SDIM can simulate an attention pattern that very similar as the target attention, therefore SDIM can match or even surpass the performance of DIN(Long Seq.).

(2) SDIM performs better than all baseline models that are proposed for modeling long-term user behaviors. Concretely, SDIM outperforms SIM by 1.62\%, outperforms UBR4CTR by 1.02\%, and outperforms ETA by 1.01\%. We also notice that DIN(Long Seq.) brings 2.21\% improvement on AUC compared to DIN, which indicates the importance of modeling the long-term user behaviors for CTR prediction. SIM, UBR4CTR and ETA performs worse than DIN(Long Seq.), which is due to the loss of information caused by the retrieval of user behaviors. These retrieval operations may be helpful in removing noise away from the sequence, but is harmful when there is not enough informative behavior for top-$k$ retrieval.

(3) SDIM is much more efficient than DIN(Long Seq.), SIM and UBR4CTR. The efficiency improvement can be ascribed to reducing the time complexity of the algorithm to $O(Lm\log(d))$.
SDIM is also much more efficient than ETA \cite{DBLP:journals/corr/abs-2108-04468}. ETA also uses LSH to hash the target item and behavior sequence, the time complexity of the hash operation is the same as SDIM. After hashing, ETA calculates the hamming distance and selects top-$k$ items for target attention, the time complexity is $O(BL+Bkd)$. While SDIM only introduces a gather operator followed by a mean pooling of $m/ \tau$ hashes, thus is much more efficient than ETA.

\subsection{Results on Industrial Dataset}

\begin{figure}[tb]
	\centering 
	\includegraphics[width=0.85\linewidth]{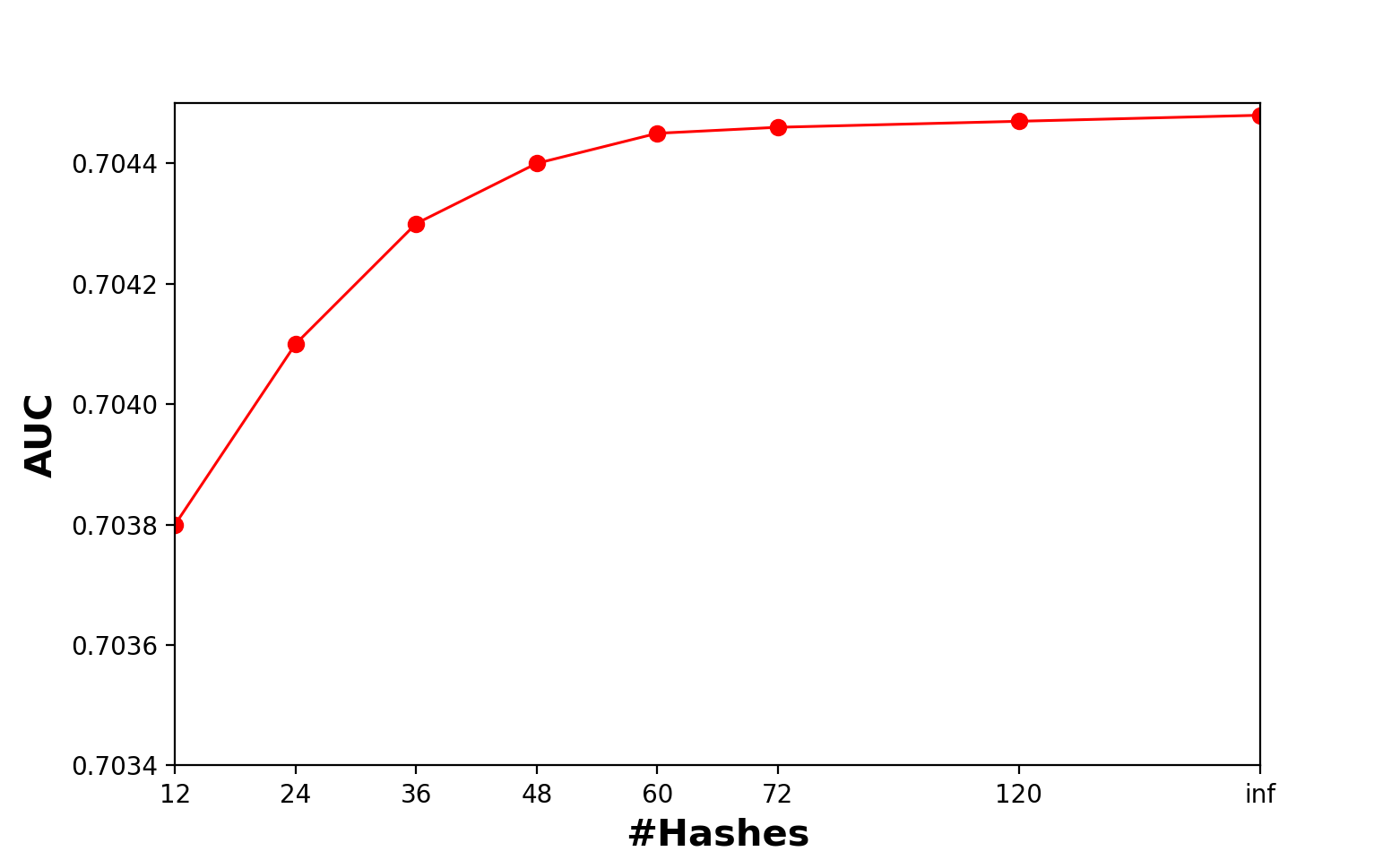}  
	\caption{AUC results of SDIM on industrial dataset when altering the number of hashes $m$.}  
	\label{fig5}   
\end{figure}

The overall results of different models on industrial dataset are shown in Table~\ref{result:2}. Similar to the results from Taobao dataset, SDIM outperforms all competitive baselines, and performs on par with DIN(Long Seq.). Our SDIM achieves 1.92\%, 2.08\%, 1.38\% improvements compared with SIM, UBR4CTR, and ETA respectively, while being much faster than these methods.

Since the user sequence length in industrial Dataset is large enough which is friendly to retrieve-based methods, it seems that they should perform on par with DIN(Long Seq.). However, the results in Table~\ref{result:2} show that their performance has some gaps with DIN(Long Seq.).
We argue that this is because user's interest are typically diverse and people often want to buy new categories of goods, especially in our food search cases. When faced with a candidate item in a new category, it's difficult for these retrieve algorithms to accurately pick out the most valuable items from users' historical behaviors. 

Compared with the Taobao dataset, the industrial dataset contains more candidate items per request, so SDIM can achieve more speedups on this dataset. The industrial dataset also has longer user behavior sequences (T=1024), so retrieve-based methods also achieve more speedups. The experimental results demonstrate the superiority of SDIM.

\subsection{Hyper-Parameter Analysis}
There are two important hyper-parameters in SDIM: (1) the number of hashes $m$, and (2) the width of hash signatures $\tau$.

\subsubsection{Analysis on $m$}
\begin{table}[t!]
\begin{center}
\setlength{\tabcolsep}{6.0mm}{
\begin{tabular}{l|c}
\toprule \bf Hyper-Parameter & \bf AUC \\ \midrule
{$\tau=1$} & 0.6911 \\ 
{$\tau=2$} & 0.7032 \\
{$\tau=3$} & \bf 0.7044 \\ 
{$\tau=5$} & 0.7034 \\
{$\tau=10$} & 0.6923 \\ 
\bottomrule
\end{tabular}}
\end{center}
\caption{\label{result:3} AUC results of SDIM on industrial dataset when altering the width parameter $\tau$.}
\end{table}

The number of hashes $m$ affects the estimation error of the proposed hashing-based attention. As the number of sampled hash function increases, the estimated user interest will be more close to $\mathbb{E}\left[\text{Attn}(\mathbf{q}, \mathbf{S}) \right]$ in Eq.~\ref{EAttn}. 

To assess the estimation error, 
we test the performance of SDIM using $m$ hashes, where $m \in \{24,36,48,60,72,90,120\}$. We also implement a variation of SDIM that directly compute attention weights using the expected collision probability $\mathbb{E}\left[\tilde{p}_{j}\right]$ in Eq.~\ref{epj}. This baseline simulates the behavior of SDIM when the number of hash signatures tends to be infinite. The results are shown in Figure~\ref{fig5}.
It can be seen that when $m>48$ the models perform almost the same. For the sake of efficiency, we use $m=48$ in our model.

\subsubsection{Analysis on $\tau$}
As we described in subsection~\ref{sec-4.2.2}, $\tau$ controls the strength of the model on paying more attention to more similar items. We investigate different attention patterns of SDIM by varying $\tau$ in \{1,2,3,5,10\}. The results are shown in Table~\ref{result:3}.

From Table~\ref{result:3}, we can see that SDIM performs well when $2 \leq \tau \leq 5$. For the balance of effectiveness and efficiency, we use $\tau=3$ in our online model. The model performs poorly when $\tau=1$ because it encodes too many noise behaviors. On the contrary, the model performs poorly when $\tau=10$ as only very similar items have the chance to contribute to user interest, which is unfriendly to users with few behaviors.

\begin{table}[t!]
\begin{center}
\setlength{\tabcolsep}{3.7mm}{
\begin{tabular}{l|ccc}
\toprule \bf Method & \bf AUC & \bf T\&I Speed \\  \midrule
{DIN(T=16)} & 0.8627 & - \\ 
{SDIM(T=16)} & 0.8637 & 2.0x \\ \bottomrule
\end{tabular}}
\end{center}
\caption{\label{result:5} Performance comparison of different model on modeling short-term user's interests.}
\end{table}

\subsection{Experiments on Short-Term User Behavior Modeling}
We also conduct an extra experiment to test SDIM's performance on modeling short-term user behaviors. But please note that SDIM is mainly proposed to solve the problem of long-term user interest modeling for industrial recommendation systems, and one can directly plug in the full target attention or more complexity module to model short sequence. We conduct this experiments to just show model's performance on special cases. We conduct this experiment on Taobao dataset, and the results are shown in Table~\ref{result:5}.

The results demonstrate that SDIM can still achieve comparable results with standard target attention model on modeling short sequence, while being more efficient.

\subsection{Online A/B Test} 
\begin{table}[t!]
\begin{center}
\setlength{\tabcolsep}{3.5mm}{
\begin{tabular}{l|cccc}
\toprule \bf Method & \bf CTR & \bf VBR & \bf Inf. Time \\  \midrule
{Base(w/o long seq.)} & - & - & $\approx$ 60ms \\ \hline
{DIN(T=1024)} & \multicolumn{3}{c}{can not deploy} \\ 
{SDIM(T=1024)} & +2.39\% & +2.21\% & +1ms \\
{SDIM(T=2000)} & +2.98\% & +2.69\% & +1ms \\ \bottomrule
\end{tabular}}
\end{center}
\caption{\label{result:ab} Online A/B testing results.}
\end{table}

A strict online A/B testing is also conducted to verify the effectiveness of SDIM. For online A/B testing, the baseline model is the previous online CTR model in Meituan search system, where only short-term user behavior sequences are used. 
The test model adopts the same structure and features as the base model, but incorporates a long-term user interests modeling module with users recent 1,024 or 2,000 behaviors on this basis. We use the proposed SDIM to model long-term user interests, and denote this test model as SDIM(T=1024) and SDIM(T=2000). 
The testing lasts for 14 days, with 10\% of Meituan search traffic are distributed for each model respectively. The results of A/B testing are shown in Table~\ref{result:ab}.

SDIM(T=2000) achieves 2.98\% improvements ($p\text{-value} < 0.001$) on CTR and 2.69\% improvements ($p\text{-value} < 0.005$) on VBR compared with the base model, which can greatly increase online profit considering the large traffic of Meituan APP. 
The inference time of SDIM(T=2000) increased by 1ms compared with the Base(w/o long seq.). The increase in inference time is mostly caused by the transmission time between BSE Server and CTR Server.

We also tried to deploy the model that directly uses the target attention to model long-term user behavior sequences with $T=1024$. However, its inference time is greatly increased by about 50\% (25ms-30ms), which is unacceptable for our system. Therefore we can't leave this model online for 14 days for A/B Testing. SDIM performs on par with this model, but saves 95\% online inference time. SDIM has now been deployed online and serve the main traffic of Meituan's home search system.

\section{Conclusions}
In this paper, we propose a hash sampling-based method named SDIM for modeling long-term user behaviors. Instead of designing complicated modules to retrieve from long-term user behaviors, we directly gather behavior items associated with the candidate item with the same hash signature to form the user interest.
We also propose a new online serving framework that decouples the hashing of user behavior sequence from the entire model, which makes it latency-free for the CTR server.
We show that the proposed method performs on par with DIN(Long Seq.), while being sizable times faster. SDIM has been deployed online in Meituan APP.

Future work includes reducing the transmission cost, exploring more complex structures such as multi-head hashing, and so on.

\bibliographystyle{ACM-Reference-Format}
\bibliography{ref}

\end{document}